\documentclass[a4paper,fleqn,usenatbib]{mnras}
\usepackage{newtxtext,newtxmath}
\usepackage[T1]{fontenc}
\usepackage{ae,aecompl}

\usepackage{graphicx}	% Including figure files
\usepackage{amsmath}	% Advanced maths commands
\usepackage{amssymb}	% Extra maths symbols
\usepackage{multirow}
\usepackage{color}
\usepackage{hyperref}

\title[UX~Ori variability in CO~Ori]{Probing the origin of UX~Ori-type variability in the YSO binary CO~Ori with VLTI/GRAVITY\thanks{Based on observations collected at the European Southern Observatory, Chile under program ID: 60.A-9159(A)}
}

\author[C. L. Davies et al.]{C.~L. Davies$^{1}$\thanks{E-mail: cdavies@astro.ex.ac.uk},  A. Kreplin$^{1}$, J. Kluska$^{1}$, E. Hone$^{1}$, S. Kraus$^{1}$
\\
$^{1}$School of Physics, University of Exeter, Physics Building, Stocker Road, Exeter EX4 4QL, UK\\
}

\date{Accepted 2017 November 24. Received 2017 October 18; in original form 2017 August 18}

\pubyear{2017}

\begin{document}
\label{firstpage}
\pagerange{\pageref{firstpage}--\pageref{lastpage}}
\maketitle

\begin{abstract}
The primary star in the young stellar object (YSO) binary CO~Ori displays UX~Ori-type variability: irregular, high amplitude optical and near-infrared photometric fluctuations where flux minima coincide with polarization maxima. This is attributed to changes in local opacity. In CO~Ori~A, these variations exhibit a $12.4\,$yr cycle. Here, we investigate the physical origin of the fluctuating opacity and its periodicity using interferometric observations of CO~Ori obtained using VLTI/GRAVITY. Continuum $K$-band circum-primary and circum-secondary emission are marginally spatially resolved for the first time while Br$\gamma$ emission is detected in the spectrum of the secondary. We estimate a spectral type range for CO~Ori~B of K2-K5 assuming visual extinction, $A_{\rm{V}}=2$ and a distance of $430\,$pc. From geometric modelling of the continuum visibilities, the circum-primary emission is consistent with a central point source plus a Gaussian component with a full-width-half-maximum of $2.31\pm0.04\,$milliarcseconds (mas), inclined at $30.2\pm2.2^{\circ}$ and with a major axis position angle of $40\pm6^{\circ}$. This inclination is lower than that reported for the discs of other UX~Ori-type stars, providing a first indication that the UX~Ori phenomena may arise through fluctuations in circum-stellar material exterior to a disc, e.g. in a dusty outflow. An additional wide, symmetric Gaussian component is required to fit the visibilities of CO~Ori~B, signifying a contribution from scattered light. Finally, closure phases of CO~Ori~A were used to investigate whether the $12.4\,$yr periodicity is associated with an undetected third component, as has been previously suggested. We rule out any additional companions contributing more than $3.6\%$ to the $K$-band flux within $\sim7.3-20\,$mas of CO~Ori~A.
\end{abstract}


%%%%%%%%%%%%%%%%%%%%%%%%%%%%%%%%%%%%%%%%%%%%%%%%%%

%%%%%%%%%%%%%%%%% BODY OF PAPER %%%%%%%%%%%%%%%%%%

\section{Introduction}
The T~Tauri primary star CO~Ori~A (spectral type G0; \citealt{Calvet04}), located in the $6-7\,$Myr old $\lambda$~Orionis region \citep{Dolan01} displays strong levels of irregular photometric variability across optical and infrared (IR) wavelengths \citep[e.g.][]{Shenavrin11, Shenavrin16} leading to its identification as a UX~Ori-type variable star. In addition to their high-amplitude variability ($\Delta V\approx2-3\,$mag), UX~Ori stars exhibit high linear polarization during their periods of minimum brightness \citep[e.g.][]{Grinin91, Rostopchina07}. The origin for the variability is uncertain. \citet{Herbst99} argue the phenomena arises due to unstable accretion onto these stars while an origin in the disc \citep{Grinin91, Natta97} or a dusty outflow \citep{Vinkovic07, Tambovtseva08} have also been suggested. 

Evidence has been growing in support of a disc-based origin as previous geometric modelling of high resolution infrared and sub-millimetre interferometric observations of UX~Ori-type stars have begun to reveal a preference towards intermediate to high disc inclinations \citep{Eisner04, Pontoppidan07, Chapillon08, Kreplin13, Vural14, Kreplin16}. For CO~Ori in particular, analysis of optical photometry over temporal baselines of up to $100$\,yrs has revealed a long-term periodic cycle of $\sim11-12\,$yrs \citep{Grinin98, Rostopchina07}. This periodicity is accompanied by a polarization cycle of similar period in which photometric minima occur during polarization maxima \citep{Rostopchina07}. The timescale of the periodicity is much longer than would be expected if the variability arose from Keplerian rotation of orbiting material close to the accretion region ($\sim$days). Instead, it suggests the presence of long-lived asymmetric features in the circumstellar material which could be associated with e.g. ongoing planet formation or the presence of an additional low mass companion in the CO~Ori system which may perturb the circum-primary disc \citep{Rostopchina07, Demidova10}. 

By comparison, the secondary, CO~Ori~B, is not so well-studied. The object was first reported as an additional emission component centred $\sim2$'' from CO~Ori at a position angle of $\sim280^{\circ}$ \citep{Reipurth93} and has since been confirmed through Variability-Induced Mover modelling of the Hipparcos photo-centre displacement of CO~Ori \citep{Bertout99} and adaptive optics imaging with VLT/NACO \citep{Correia06}. At the newly-determined GAIA distance to the CO~Ori system of $430\pm90\,$pc \citep{Lindegren16}, this corresponds to a physical separation of $\sim860\,$au. The CO~Ori system has been studied as part of the ongoing ``Multiplicity of Herbig Ae/Be Stars'' project (c.f. \citealt{Bliek14}). These authors have confirmed the binary status of CO~Ori~B based on (i) the local surface density in the $\lambda$ Orionis region which suggests a statistical probability of $>99.99\%$ that the two stars, separated as they are by $2.0$'', are physically related and (ii) observations from 1993 and 2005 confirm the common proper motion of the pair (\citealt{Bliek14}; B.\ Rodgers, private communication). 

Contributing less than $10\%$ to the optical flux of the CO~Ori system \citep{Rostopchina07}, CO~Ori~B is unlikely to contribute to the observed UX~Ori-type variability. However, photometric monitoring at optical wavelengths has shown that CO~Ori~B is also variable on the order of days \citep{Pierpoint03}. In addition, the presence of Br$\gamma$ and Pa$\beta$ emission in its spectrum \citep{Rodgers03}, typically associated with the high-energy accretion process in young stellar objects (YSOs), indicates the likely existence of a circum-secondary disc. 

Here, we report on the first marginally resolved detection of circumstellar $K$-band emission from both components of the CO~Ori binary system using infrared long-baseline interferometry. Details of our observations are presented in Section~\ref{sec:obs}. Assuming that CO~Ori~B contributes $10\%$ to the optical flux, we use optical photometry and stellar parameters taken from \citet{Calvet04} to estimate a likely spectral type range for the secondary. Combining this with $K$-band flux measurements for both objects, we used spectral energy distribution (SED) fitting to calculate the stellar flux contribution to our observed $K$-band emission. The observed continuum visibilities were used to characterise the circum-primary and circum-secondary emission and constrain the viewing geometry of the circum-primary disc. The modelling process and results are presented in Section~\ref{sec:disc}. We also used the closure phases to investigate whether the periodicity observed in the photometry and polarization of CO~Ori~A could be associated with a hitherto unknown additional companion. The results of this analysis are presented in Section~\ref{sec:rnd}. We summarise our conclusions in Section~\ref{sec:concl}.

\section{Observations: VLTI/GRAVITY Spectro-interferometry} \label{sec:obs}
\begin{table}
 \centering
 \caption{CO~Ori Observation Log}
 \label{tab:obslog}
 \begin{tabular}{clcc}
    \hline
    Target & UT Date & Stations & Calibrator(s)\\
    \hline
    CO~Ori~A & 2016 Sep 11 & A0-B2-C1-D0 & HD~34471\\
    \multirow{2}{*}{CO~Ori~B} & \multirow{2}{*}{2016 Sep 12} & \multirow{2}{*}{A0-B2-C1-D0} & HD~34471\\
    & & & HD~38117\\
    \hline
 \end{tabular}
\end{table}

\begin{figure*}
  \centering
   \includegraphics[width=0.7\textwidth]{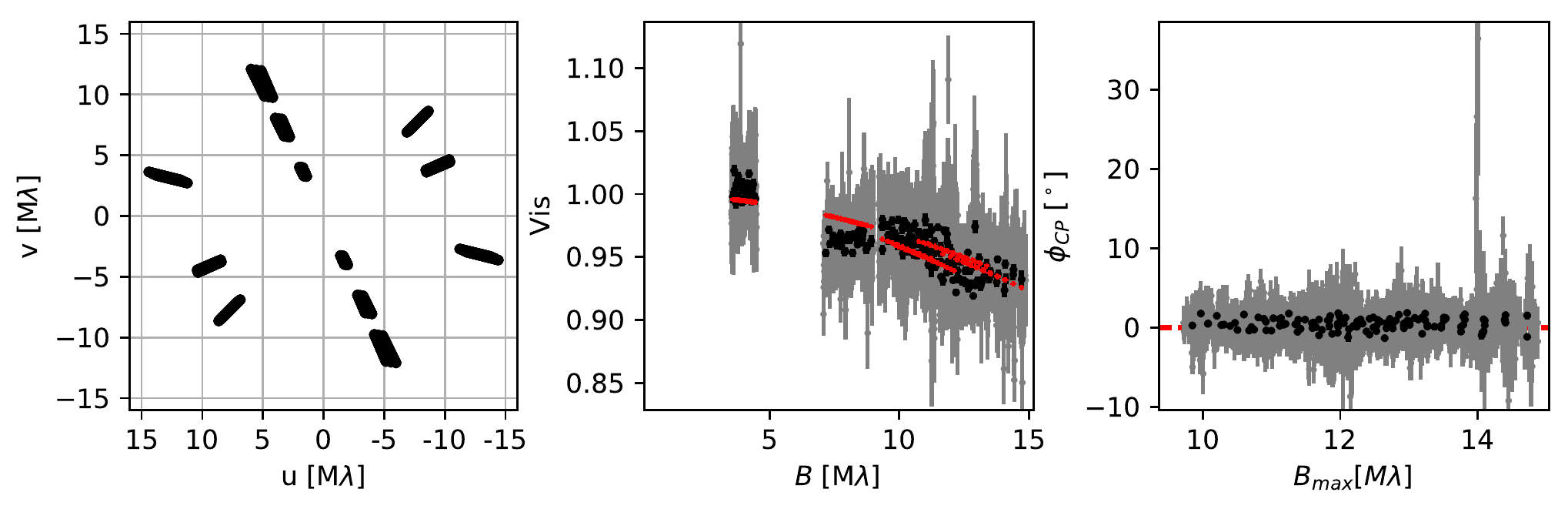}
   \includegraphics[width=0.7\textwidth]{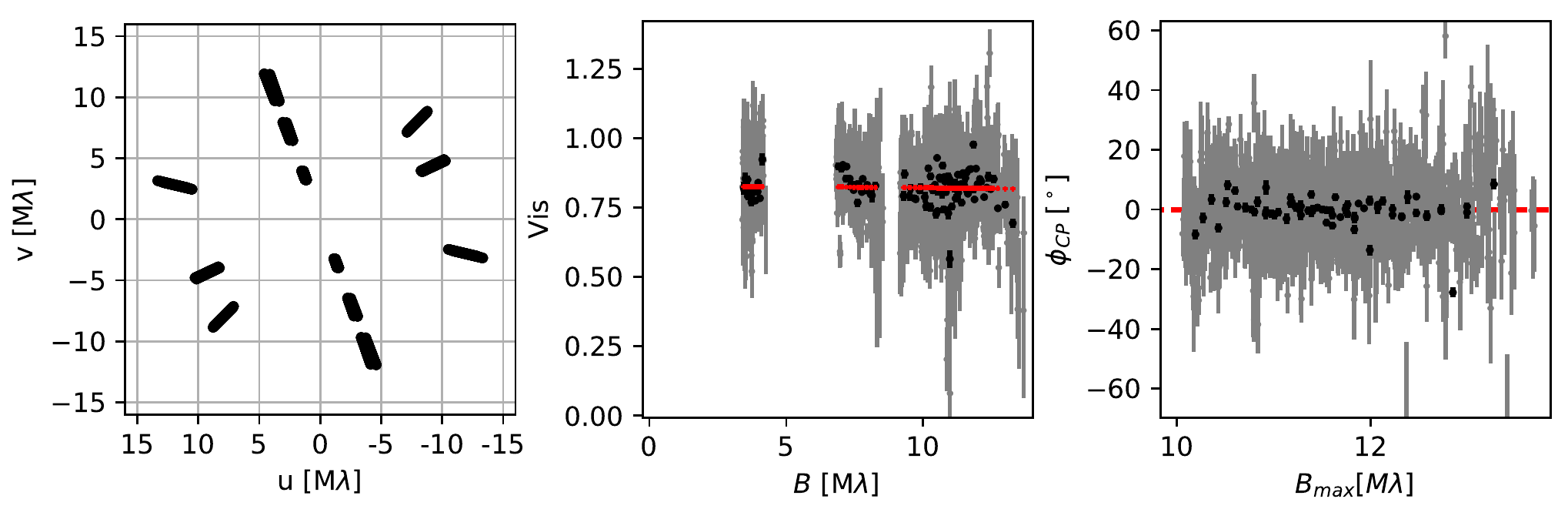}
   \caption{Left panels: ($u, v$)-plane coverage of our VLTI/GRAVITY single-field observations of CO~Ori~A (top) and dual-field observations of CO~Ori~B (bottom). In both cases, north is up; east is left. Middle and right panels: visibility amplitudes and closure phases for CO~Ori~A (top) and CO~Ori~B (bottom). The grey points are the full data set while black points are the binned data used in our analysis (see text for details). The red points in the middle panels correspond to the models in Table~\ref{tab:fits} while the red line in the right panels represents the closure phase signal expected for a centro-symmetric model ($\phi_{\rm{CP}}=0^{\circ}$).
  \label{fig:obsQuan}}
\end{figure*}

We observed the CO~Ori binary system with the Very Large Telescope Interferometer (VLTI) four-telescope $K$-band beam combiner, GRAVITY \citep{Eisenhauer11, Gravity17} on the Auxiliary Telescopes (ATs) during the September 2016 period of Science Verification. An overview of the observations is presented in Table~\ref{tab:obslog}. Each target was observed once: CO~Ori~A in single field fringe-tracking mode and CO~Ori~B in dual field field-tracking mode. In the former, light from CO~Ori~A was split with 50$\%$ being fed into the fringe tracker and the other 50$\%$ to the science detector. For the dual field observations, the light from CO~Ori~A was used for fringe tracking while 100\% of the $K$-band light from CO~Ori~B was fed into the science camera. This procedure allows for longer integration times on the fainter science target as tracking and correcting for atmospheric turbulence is undertaken in real-time using the brighter fringe-tracker star. This enables one to obtain fringes on objects down to $\sim10\,$mag in $K$-band on the ATs of the VLTI. The ($u,v$)-plane coverage for the observations of the primary (top) and secondary (bottom) component of CO~Ori are shown in the left panels of Fig.~\ref{fig:obsQuan}. 

Both the science and the fringe tracker data were available for analysis: our science detector data were spectrally dispersed at medium resolution ($\Delta R/R=500$ from $1.99$ to $2.45\,\mu$m) while fringe tracker data products are always dispersed at low resolution ($\Delta R/R=22$). The light was combined from the four ATs in the compact array configuration (A0-B2-C1-D0). This configuration provided projected baseline lengths, $B$, between $9$ and $30\,$m, corresponding to a maximum angular resolution of $\lambda/2B=7.3\,$milliarcseconds (mas), where $\lambda$ denotes the wavelength of our observations ($2.1\,\mu$m). This enabled us to probe regions down to $\sim3.1\,$au from the stars. 

\begin{figure}
  \centering
   \includegraphics[width=\columnwidth]{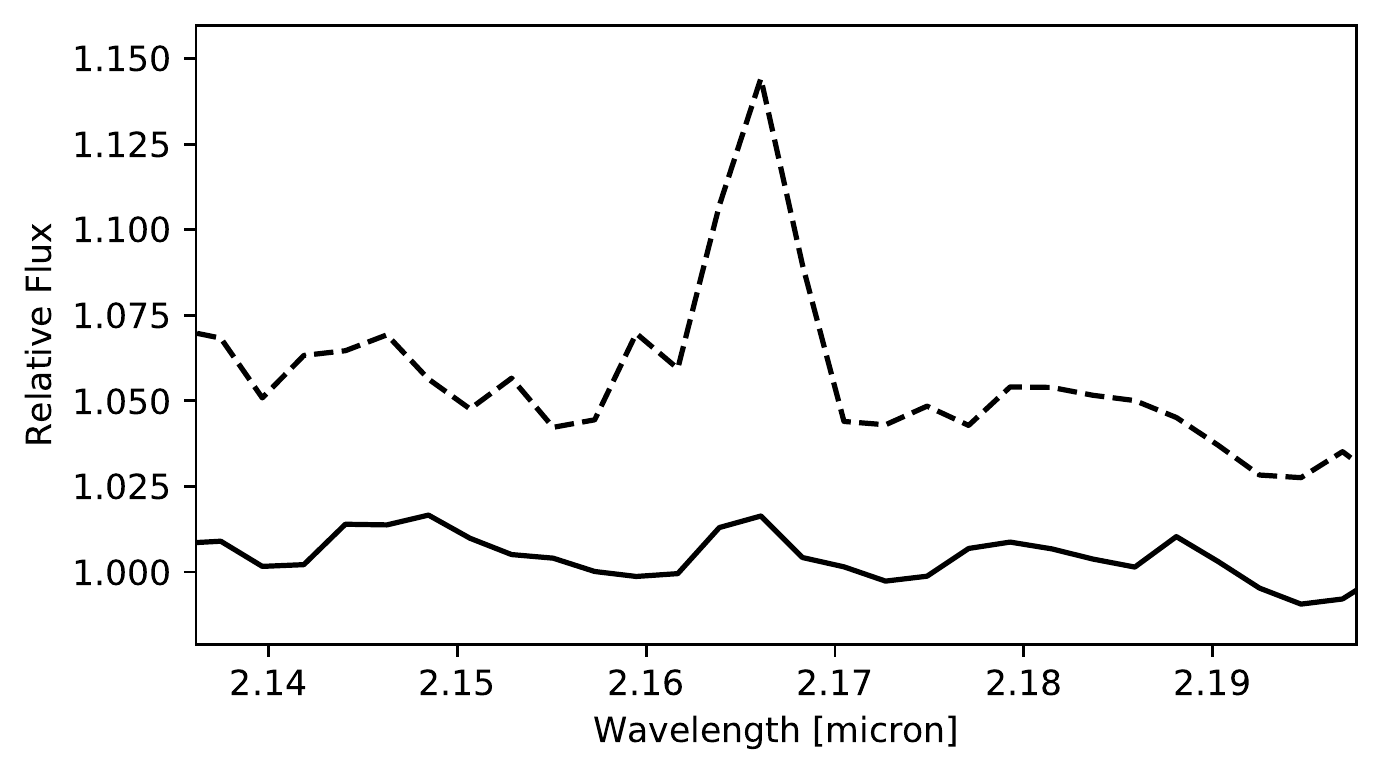}
   \caption{Normalised $K$-band spectra of CO~Ori~A (solid line) and CO~Ori~B (dashed line), focusing on the Br$\gamma$ emitting region at $2.166\,\mu$m. The spectra have been offset with respect to each other for clarity. CO\,Ori~B displays Br$\gamma$ emission at a strength of $\sim10\%$ above continuum.\label{fig:BrGam}}
\end{figure}

The data were reduced using the standard GRAVITY reduction pipeline (Version 0.8.4; \citealt{Lapeyrere14}) using the default parameters. In particular, we used the so-called ``v-factor'' which estimates the loss of coherence in the science detector using the fringe tracker data and rescales the science absolute visibilities accordingly. Calibrator stars HD~34471 and HD~38117 were observed based on their spectral types. As K0 stars, Br$\gamma$ absorption features typically seen in the spectra of earlier type stars should be absent. This was confirmed via visual inspection of the calibrator spectra. The medium resolution spectra for CO~Ori~A (lower solid line) and CO~Ori~B (upper dashed line), normalised by the calibrator spectra and centered on the Br$\gamma$ feature at $2.166\,\mu$m, are shown in Fig.~\ref{fig:BrGam}. As previously reported by \citet{Rodgers03}, we see that CO~Ori~B is responsible for the majority of the Br$\gamma$ emission observed in the system with a line strength $\sim10\%$ above the continuum. It was not possible to distinguish a signal from the noise in the spectrum of CO~Ori~A. Using the normalised spectra, we attempted to extract differential phases and visibilities across the Br$\gamma$ feature of CO~Ori~B. However, no changes across the spectral line along any of the baseline position angles probed were detected. This suggests that the line-emitting region is coincident with the continuum-emitting region as a line-emitting region more (less) extended than the continuum would result in a decrease (increase) in visibility and a phase shift across the line. Henceforth, our analysis only concerns the continuum visibilities and closure phases.

HD~34471 and HD~38117 were used to calibrate the measured squared visibility amplitudes and closure phases. Their uniform disc (UD) diameters were acquired from JMMC SearchCal \citep{Bonneau06, Bonneau11}. The transfer functions for both observation nights appeared stable. In the middle and right panels of Fig.~\ref{fig:obsQuan}, we show the resulting calibrated continuum visibilities and closure phases for CO~Ori~A (top) and CO~Ori~B (bottom), respectively. The gray points represent the full medium spectral resolution data set while the data represented by the black points have been binned to increase the signal-to-noise. The size of the bins is $20$ wavelength channels. The binned data set for CO~Ori~A provides similar signal-to-noise as the low resolution, fringe tracker data obtained during the single field observations but provides twice the spectral resolution. We provide a comparison between the fringe tracker and science target data in Appendix~\ref{app:FTvsSC}. We used the binned medium resolution data set in the analysis we present in the following sections.

\section{Visibility modelling}\label{sec:disc}
Overlaid on the closure phase data in the right panel of Fig.~\ref{fig:obsQuan} is a dashed horizontal red line corresponding to the closure phases expected for a centro-symmetric brightness distribution: $\phi_{\rm{CP}}=0^{\circ}$. Both objects in the CO~Ori binary system are consistent with having $\phi_{\rm{CP}}=0^{\circ}$ at all baseline lengths probed. As such, we only consider centrally symmetric brightness distribution profiles when modelling the geometry of the circum-primary and circum-secondary emission.

As can be seen from the middle panel of Fig.~\ref{fig:obsQuan}, both components of the CO~Ori binary system appear marginally resolved. CO~Ori~A displays a slight drop in visibility with increasing baseline length while the visibilities of CO~Ori~B appear roughly flat at $\sim0.8$. The observed visibilities of both targets comprise both stellar and circumstellar emission components. The stellar emission is expected to be unresolved in both cases as the stellar radii are much smaller than our resolution limit of $\sim7.3\,$mas ($\sim3.1\,$au). Therefore, we fixed the stellar contribution to the visibility, $V_{\star}=1$. The circumstellar visibility, $V_{\rm{CS}}$, then becomes,
\begin{equation}
V_{\rm{CS}} = \frac{|V_{\rm{obs}}(F_{\star}+F_{\rm{CS}}) - V_{\star} F_{\star}|}{F_{\rm{CS}}}.
\end{equation}
Here, $V_{\rm{obs}}$ is the observed visibility and $F_{\star}$ and $F_{\rm{CS}}$ are the stellar and circumstellar flux contributions, respectively. 

\begin{table}
 \centering
 \caption{Parameters adopted in SED fitting. Column~1: binary component name; column~2: $V$-band magnitudes of each component, determined assuming a total $V$-band magnitude for the CO~Ori system of $11.58$ \citep{Calvet04} and a secondary flux contribution of $10\%$; column~3: $K$-band magnitudes of CO~Ori~A and CO~Ori~B, taken from \citet{Calvet04} and \citet{Rodgers03}, respectively; column~4: $V$-band extinction, $A_{\rm{V}}$, from \citet{Calvet04}; column~5: spectral types (SpT) taken from \citet{Calvet04} for CO~Ori~A and estimated for CO~Ori~B from the difference in $V$-band magnitude; column~6: adopted stellar effective temperature based on the spectral type.}
 \label{tab:SEDfit}
 \begin{tabular}{cccccc}
 \hline
 Target & $V$ & $K$ & $A_{\rm{V}}$ & SpT & $T_{\rm{eff}}$\\
  & (mag) & (mag) & & (K) & \\
 \hline
 CO~Ori~A & 11.69 & 6.72 & 2.0 & G0& 6030 \\
 CO~Ori~B & 14.08 & 9.50 & 2.0 & K2-K5& 4500\\
 \hline
 \end{tabular}
\end{table}

Values of $F_{\star}$ in specific wavebands are typically estimated via SED fitting. For this, stellar photosphere-tracing photometry and an estimate of the spectral type or effective temperature for each star are required. For CO~Ori~A, we retrieved photometry and stellar parameters from \citet{Calvet04} and present these in Table~\ref{tab:SEDfit}. However, no published optical photometry or spectral type was available in the literature for CO~Ori~B. Instead, we use the $10\%$ upper limit to the optical flux contribution reported in \citet{Rostopchina07} to estimate the individual contributions to the $V$-band magnitude of $11.58$ reported in \citet[][see Table~\ref{tab:SEDfit}]{Calvet04}. In addition, we estimate the spectral type range for CO~Ori~B of K2-K5 based on the assumed difference in $V$-band flux using Table~15.7 of \citet{Cox00}. This implicitly assumes that the $V$-band extinction (see Table~\ref{tab:SEDfit}) and distance modulus for CO~Ori~B are the same as for CO~Ori~A. We used the spectral type-to-effective temperature ($T_{\rm{eff}}$) scale of \citet{Pecaut13} to convert this into a $T_{\rm{eff}}$ range of $4140-4760\,$K. 

\citet{Kurucz79} model atmospheres for $T_{\rm{eff}}=6000$ and $4500\,$K were then fit to the individual de-reddened $V$-band fluxes for CO~Ori~A and CO~Ori~B, respectively. From these fits, an estimate of the primary and secondary stellar $K$-band flux was made. Finally, these values were compared to the observed $K$-band fluxes reported in \citet{Calvet04} and \citet[][see Table~\ref{tab:SEDfit}]{Rodgers05} to estimate $F_{\star}$ for each of the binary components (see Table~\ref{tab:fits}). The values of $F_{\star}$ were fixed during the visibility modelling.

We modelled the circumstellar emission component as a single two-dimensional (2D) Gaussian brightness distribution. The free parameters of such a model are the full width at half maximum (FWHM), the elongation ratio of the major and minor FWHM, and the position angle of the major axis, measured East of North. These define the characteristic radius of the emitting region and, assuming the emission arises from a circular disc, its inclination, $i$ ($i=0^{\circ}$ corresponds to a disc viewed face-on), and position angle, PA. The results of our modelling are presented in Table~\ref{tab:fits}. For CO~Ori~A, the Gaussian-plus-point source (G+PS) model resulted in a reduced-$\chi^{2}$ ($\chi^{2}_{\rm{r}}$) of $8.04$. For CO~Ori~B, we included an additional halo (H) component (a symmetric 2D Gaussian with fixed FWHM of $1000\,$mas) as the visibilities at the shortest baselines suggest the presence of an over-resolved component. Due to the inferred late spectral type of the object (K2-K5), this halo component is likley associated with scattered light \citep{Pinte08}. We were unable to constrain the inclination and position angle of the Gaussian model component for CO~Ori~B and instead fixed these parameters to zero. The point source-plus-Gaussian-plus-Halo (PS+G+H) model for CO~Ori~B produced an improved fit to the data compared to the G+PS model with no $i$ or PA constraints: $\chi^{2}_{\rm{r}}=15.06$ compared to $55.35$.

The errors on each of the best-fit model parameters were estimated via bootstrapping. A total of $1000$ new realizations of our original visibility data sets were created and put through the same modelling pipeline as the original data. We ensured that the initial values of the model parameters were kept consistent throughout. Histograms were created from the resulting spread of model parameters and the errors on each parameter were estimated from a $1\sigma$-Gaussian fit to each histogram. 

\begin{table}
\centering
\caption{Results of geometric model fitting to the continuum visibilities. The subscripts H and G refer to the broad, symmetric 2D Gaussian (i.e halo) component and the non-symmetric 2D Gaussian component (i.e. disc), respectively.}
\label{tab:fits}
\begin{tabular}{lcc}
\hline
 & CO Ori A & CO Ori B
\\
\hline
$F_{\star}$ ($\%$)  & $3.9$ (fixed) & $19.7$ (fixed) \\
$F_{\rm{H}}$ ($\%$) & --            & $17.4\pm3.0$ \\
FWHM$_{\rm{H}}$ (mas) & -- & $1000$ (fixed) \\
FWHM$_{\rm{G}}$ (mas)          & $2.31\pm0.04$ & $0.96\pm0.55$ \\
$i$ ($^{\circ}$)    & $30.2\pm2.2$  & $0$ (fixed)\\
PA ($^{\circ}$)     & $40\pm6$      & $0$ (fixed)\\
\hline
$\chi^{2}_{\rm{r}}$    & $8.04$    & $15.06$\\
\hline
\end{tabular}
\end{table}

CO~Ori~A has previously been observed at $H$-band with VLTI/PIONIER \citep{leBouquin11}. Geometric model fits to the PIONIER visibilities indicate the presence of a disc inclined at $48.7^{\circ}$ with major axis position angle of $38^{\circ}$ east of north \citep{Lazareff16}. These values are broadly consistent with our best fit parameters (see Table~\ref{tab:fits}). Their half-light semi-major axis of $0.59\,$mas for the $H$-band emission is smaller than the FWHM of $2.31\pm0.04$ we derive here for the $K$-band emission. Differences between the model adopted here and in \citet{Lazareff16} prevents direct comparison between these values but the fact that the size of the $H$-band emitting region is smaller than the $K$-band emitting region is expected as shorter wavelengths likely trace hotter material closer to the star. At the distance of the CO~Ori system, both the $H$- and $K$-band emission trace material at $\sim0.5-1\,$au, exterior to the expected location of the dust sublimation region. 

Intriguingly, our inferred circum-primary disc inclination of $30.2\pm2.2^{\circ}$ is toward the lower end of the range of disc inclinations measured for other UX~Ori-type stars: $\sim30-50^{\circ}$ for CQ~Tau, \citep{Eisner04, Chapillon08}, $\sim50^{\circ}$ for V1026~Sco \citep{Vural14}, and $\sim70^{\circ}$ for VV~Ser, KK~Oph and UX~Ori \citep{Pontoppidan07, Kreplin13, Kreplin16}. The intermediate to high disc inclinations of these other UX~Ori stars provided support in favour of UX~Ori-type variations being caused by dust obscuration in discs viewed at intermediate to high inclinations. In contrast, our inferred inclination for CO~Ori~A suggests that changes in the line of sight opacity above or below the disc may be required. A disc-based origin would likely only be possible if the changes in line-of-sight opacity occurred in the upper regions of a disc which is either extremely flared or exhibits a non-azimuthally symmetric scale height. The UX~Ori-type phenomena observed for CO~Ori~A could instead be associated with irregularities in a dusty outflow, for instance \citep{Vinkovic07, Tambovtseva08}, possibly akin to the centrifugally-driven disk wind proposed by \citet{Bans12}. Alternatively, the periodicity observed in the polarization and photometric variability of CO~Ori~A may indicate that the UX~Ori-type behaviour is associated with the presence of an additional low mass companion on an orbit inclined to the circum-primary disc \citep{Rostopchina07, Demidova10}. We consider this latter idea further in the following section.

\section{Search for an additional companion to CO~Ori~A}\label{sec:rnd}
\begin{figure}
  \centering
   \includegraphics[width=0.35\textwidth]{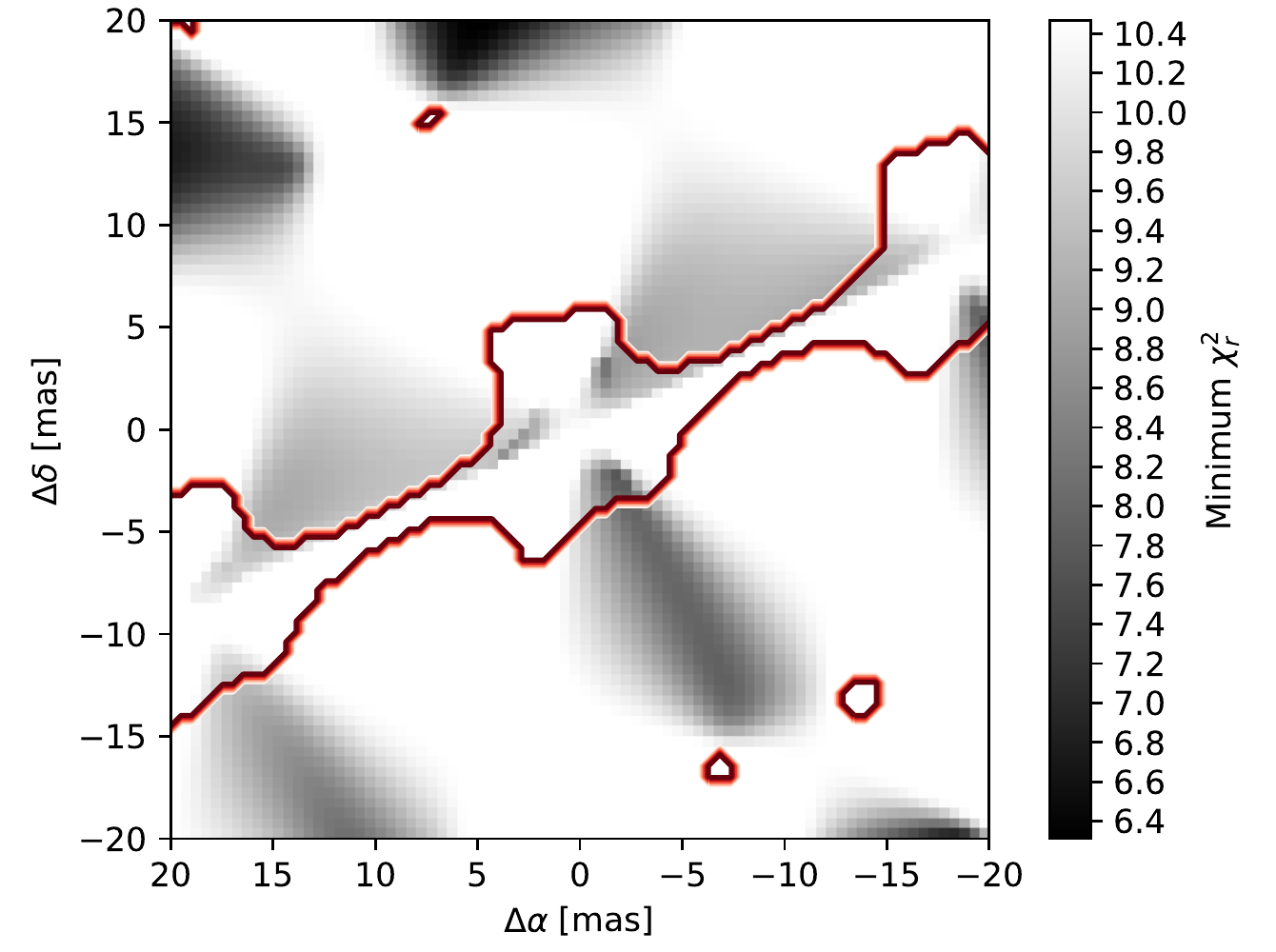}
   \includegraphics[width=0.35\textwidth]{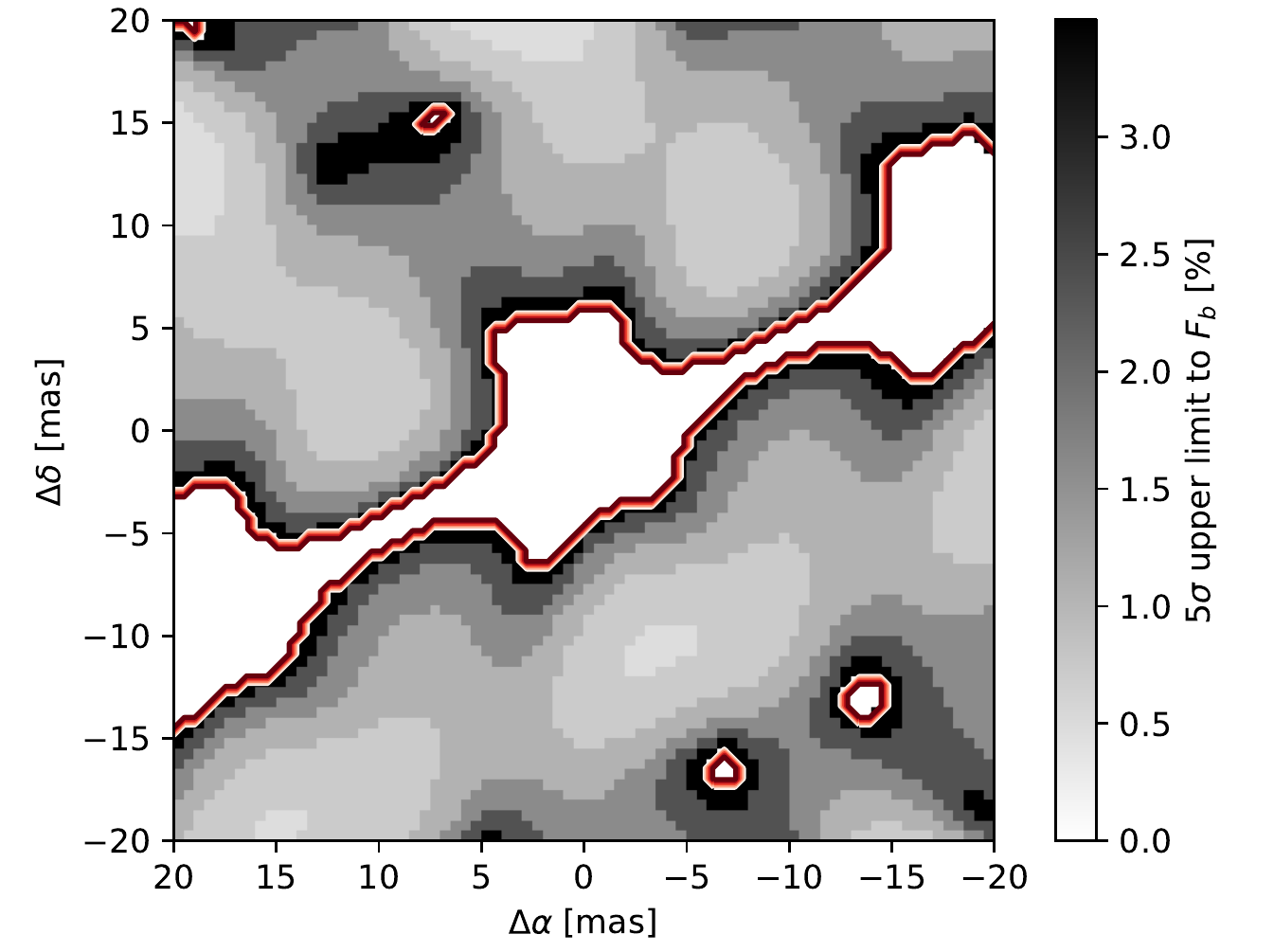}
   \caption{Top: $\chi^{2}_{\rm{r}}$ grid for the companion search around CO~Ori~A. The maximum value of $\chi^{2}_{\rm{r}}$ corresponds to that of the G+PS model for CO~Ori~A in Table~\ref{tab:fits} with a zero flux companion included (i.e. $F_{\rm{b}}=0$). Bottom: corresponding sensitivity map showing the maximum possible flux contribution of a undetected companion at each position. In both panels, the regions outlined in red mark the sky regions within which our ($u,v$)-plane coverage is not sufficient to probe.}
  \label{fig:compSearch}
\end{figure}

Fourier analysis of optical time series photometry and visual inspection of polarization measurements have revealed the presence of periodicity in the variability of CO~Ori~A. With a temporal baseline of $25\,$years, \citet{Grinin98} identified an $11.2\,$yr cycle in the $V$-band photometry and, with a temporal baseline of $100\,$yrs, \citet{Rostopchina07} identified a $12.4\,$yr periodicity in the $V$- and $B$-band photometry. \citet{Rostopchina07} also uncovered a similar period in the polarization data with polarization maxima coinciding with photometric minima. 

The presence of a periodic signal in the photometric and polarization variability indicates the possible presence of a long-lived region of enhanced opacity within the circum-primary environment. Our inferred disc inclination of $30.2\pm2.2$ for CO~Ori~A suggests that this region lies exterior to the disc or, at the very least, in the uppermost reaches of the disc around CO~Ori~A. One possible solution is that a low mass and hitherto undetected additional component to the CO~Ori system exists. Based on the results of our continuum visibility fitting (Table~\ref{tab:fits}) and adopting a stellar mass of $2.5\,\rm{M_{\odot}}$ for CO~Ori~A \citep{Calvet04}, the $12.4\,$yr periodicity could signify the presence of a companion on a Keplerian orbit at $\sim5-7\,$au. Any such companion would not have been detectable in previous multiplicity surveys of the region \citep[e.g.][]{Correia06}. With this in mind, we use the observed closure phases for CO~Ori~A (see top right panel of Fig.~\ref{fig:obsQuan}) to conduct a companion search.

We define a search grid of $40$ square mas, centered on CO~Ori~A, with a pixel resolution of $0.5\,$mas. We produce a series of models incorporating a G+PS using the FWHM, $i$, and PA determined from our visibility fitting (Table~\ref{tab:fits}) and with an additional point source component located at each possible location ($x$, $y$) in the grid. The relative contributions of the Gaussian emission and the point source are also maintained. Using the continuum closure phases, we perform a $\chi^{2}$ fit to find the best estimate of the companion flux, $F_{\rm{b}}$, at each position. The value of $\chi^{2}$, minimised for $F_{\rm{b}}$, is then divided by the number of degrees of freedom to produce the reduced-$\chi^{2}$ ($\chi^{2}_{\rm{r}}$) grid shown in the top panel of Fig.~\ref{fig:compSearch}. The maximum value in the grey-scale color bar corresponds to the $\chi^{2}_{\rm{r}}$ of the model with $F_{\rm{b}}=0$, henceforth noted as $\chi^{2}_{\rm{null,r}}$. The regions in the plot that are outlined in red mark the sky regions within which we our unable to probe due to the combined effects of our resolution limit of $\sim7.3\,$mas and our limited ($u,v$)-plane coverage. 

Following \citet{Absil11}, the probability, $P_{\rm{0}}$, for a random value of $\chi^{2}_{\rm{r}}$ in the distribution of $\chi^{2}_{\rm{r}}$  with $\nu$ degrees of freedom to equal or exceed that associated with the G+PS model with $F_{\rm{b}}=0$ is
\begin{equation}
    P_{0}= 1 - \rm{{CDF}_{\nu}} \left(\frac{\nu \chi^{2}_{\rm{null,r}}}{\rm{min}(\chi^{2}_{\rm{r}}(x, y, F_{\rm{b}}))}\right).
\end{equation}
Here, $\rm{min}(\chi^{2}_{\rm{r}}(x, y, F_{\rm{b}}))$ denotes the minimum-$\chi^{2}_{\rm{r}}$ across the entire grid and $\rm{CDF}_{\nu}$ is the $\chi^{2}$ cumulative distribution function. If the value of $P_{0}$ is less than our $5\sigma$ threshold for detection ($5.73\times10^{-5}\%$), we can reject the null hypothesis of no additional companion and the best fit G+PS model with $F_{\rm{b}}\neq0$ can be considered significant. 

We found that the G+PS model with $F_{\rm{b}}=0$ was associated with $\chi^{2}_{\rm{null,r}}/\rm{min}(\chi^{2}_{\rm{r}}(x, y, F_{\rm{b}}))=10.47$. The corresponding value of $P_{\rm{0}}$ with $\nu=119$ is $0.0008\,\%$. Thus, the minima associated with the $\chi^{2}_{\rm{r}}$ map can \emph{not} be considered significant. 

We then followed a similar method to derive a sensitivity map of upper limits to $F_{\rm{b}}$ for each ($x$, $y$) position. The $\chi^{2}_{\rm{r}}$ map in the top panel of Fig.~\ref{fig:compSearch} was used to determine a probability cube whereby
\begin{equation}
    P_{\rm{cube}}(x,y,F_{\rm{b}}) = 1 - \rm{CDF_{\nu}}\left(\frac{\nu \chi^{2}_{\rm{r}}(x,y,F_{\rm{b}})}{\chi^{2}_{\rm{null,r}}}\right).
\end{equation}
At each $x$ and $y$, $F_{\rm{b}}$ was incrementally increased until $P_{\rm{cube}}$ dropped below $5\sigma$ significance. Each value of $F_{\rm{b}}$ is then the maximum flux a potential companion at this location could have while remaining undetectable in our observations. The resulting $5\sigma$ sensitivity map is shown in the bottom panel of Fig~\ref{fig:compSearch}. From this, we can infer an upper limit to the $K$-band flux contribution of $3.6\%$ for any possible companions between $\sim7.3-20\,$mas of CO~Ori~A although we note that our search of this region is not complete due to our limited ($u,v$)-plane coverage.

\section{Conclusions}\label{sec:concl}
We report on the detection of marginally resolved spatially extended $K$-band emission from both components of the CO~Ori binary system. The observed continuum and line emission are consistent with the existence of circum-primary and circum-secondary discs.  

Our conclusions are summarised as follows:
\begin{enumerate}
    \item We find that CO~Ori~B is the main contributor to the Br$\gamma$ emission detected towards the CO~Ori system, confirming the result of \citet{Rodgers03}. The strength of the observed Br$\gamma$ emission line is $\sim10\%$ above continuum. Typically associated with the accretion phenomena in young stellar objects, the Br$\gamma$ emission from CO~Ori~B highlights the likely existence of a circum-secondary accretion disc.
    \item Assuming the secondary contributes a maximum of $10\%$ to the $V$-band flux observed in \citet{Calvet04}, we estimate a spectral type range of K2-K5 for CO~Ori~B. Using \citet{Kurucz79} model atmospheres for stellar effective temperatures of $6000$ and $4500\,$K for CO~Ori~A and CO~Ori~B, respectively, we estimated the stellar flux contributions, $F_{\star}$, to previously observed $K$-band fluxes for each object. CO~Ori~A was found to contribute $3.9\%$ to its total $K$-band emission while a value of $19.7\%$ was found for CO~Ori~B.
    \item The closure phases obtained for both CO~Ori~A and B are consistent with centro-symmetric brightness distributions.
    \item The $K$-band continuum emission from CO~Ori~A is likely associated with a disc geometry. We find that the best-fit model to the observed visibilities incorporates a central point source and a two-dimensional (2D) Gaussian component with full-width at half-maximum, FWHM$=2.31\pm0.04\,$mas. This model resulted in a $\chi^{2}_{\rm{r}}=8.04$. At the $430\,$pc distance to the CO~Ori system, the circum-primary $K$-band emission appears to trace material at $\sim1\,$au from CO~Ori~A, exterior to the expected location of the dust sublimation region. The disc major axis position angle of $40\pm6^{\circ}$ that we infer from our modelling is consistent with that found by \citet{Lazareff16} in their modelling of $H$-band emission from the object. However, our inferred disc inclination of $30.2\pm2.2^{\circ}$ is lower than that found by those same authors. Furthermore, the inclination we infer is at the lower end of the range previously inferred from prior interferometric observations of other UX~Ori stars  ($\sim30$ to $>70^{\circ}$; \citealt{Eisner04, Pontoppidan07, Chapillon08, Kreplin13, Vural14, Kreplin16}). This suggests that the changes in opacity along the line-of-sight towards CO~Ori~A do not originate within the surrounding disc as has previously been suggested for other UX~Ori-type stars. Instead, we suggest these variations likely originate from irregularities in a dusty outflow \citep{Vinkovic07, Tambovtseva08} such as a centrifugally-driven disk wind \citep{Bans12}.
    \item We detect and marginally spatially resolve near-infrared circum-\emph{secondary} continuum emission for the first time. The lack of a differential visibility and/or phase signal across the Br$\gamma$ line in the $K$-band spectrum we obtained for the object indicates that the line-emitting region is coincident with the continuum-emitting region (i.e. also marginally spatially resolved). The model which provides the best fit to the observed visibilities consists of a 2D Gaussian with FWHM$=0.96\pm0.55\,$mas and a point source component with an additional symmetric 2D Gaussian (or ``halo'') component with fixed FWHM$=1000\,$mas ($\chi^{2}_{\rm{r}}=15.06$ compared to $55.35$ without the halo component). This halo component is responsible for the observed drop in visibilities at the shortest baselines and is likely associated with a scattered light component \citep{Pinte08}. At present, we are unable to constrain the orientation of the circum-secondary disc. Longer baseline observations are required for this.
    \item We investigated whether the presence of an $11-12\,$yr periodic modulation of the variations in the optical photometry and polarization observed towards CO~Ori~A by \citet{Grinin98} and \citet{Rostopchina07} could be explained by the presence of an additional component in the CO~Ori system close to the primary star. We found no evidence for such a companion within $\sim20\,$mas of CO~Ori~A. We produced a sensitivity map to assess the level to which we could rule out an additional companion, deriving an upper limit to the $K$-band flux contribution for any undetected companion within $\sim7.3-20\,$mas of CO~Ori~A to be $3.6\%$. However, we note that this sensitivity limit is not complete along position angles of $\sim100$-$125^{\circ}$ or $\sim280$-$305^{\circ}$ due to our limited ($u,v$)-plane coverage.
\end{enumerate}

\section*{Acknowledgements}
We thank the anonymous referee for remarks that improved the manuscript. We thank the GRAVITY consortium and the Science Verification team, which is composed of ESO employees and GRAVITY consortium members (\url{https://www.eso.org/sci/activities/vltsv/gravitysv.html}) for their assistance in the preparation, conduction and reduction of our observations. We thank Bernadette Rodgers for supplying details regarding the binary nature of the CO~Ori system. The authors acknowledge support from ERC Starting Grant ``ImagePlanetFormDiscs'' (Grant Agreement No. 639889), Marie Sklodowska-Curie CIG grant (Ref. 618910), Philip Leverhulme Prize (PLP-2013-110), and STFC Rutherford Fellowship (ST/J004030/1). This research has made use of the Jean-Marie Mariotti Center SearchCal service available at \url{http://www.jmmc.fr/searchcal_page.htm}.

Facilities: VLTI.

%%%%%%%%%%%%%%%%%%%%%%%%%%%%%%%%%%%%%%%%%%%%%%%%%%



\bibliographystyle{mnras}
\bibliography{COOri} % if your bibtex file is called example.bib

%\begin{thebibliography}{99}
%\end{thebibliography}

%%%%%%%%%%%%%%%%%%%%%%%%%%%%%%%%%%%%%%%%%%%%%%%%%%

%%%%%%%%%%%%%%%%% APPENDICES %%%%%%%%%%%%%%%%%%%%%

\appendix

\section{Fringe tracker data}\label{app:FTvsSC}
\begin{figure}
  \centering
   \includegraphics[width=0.45\textwidth]{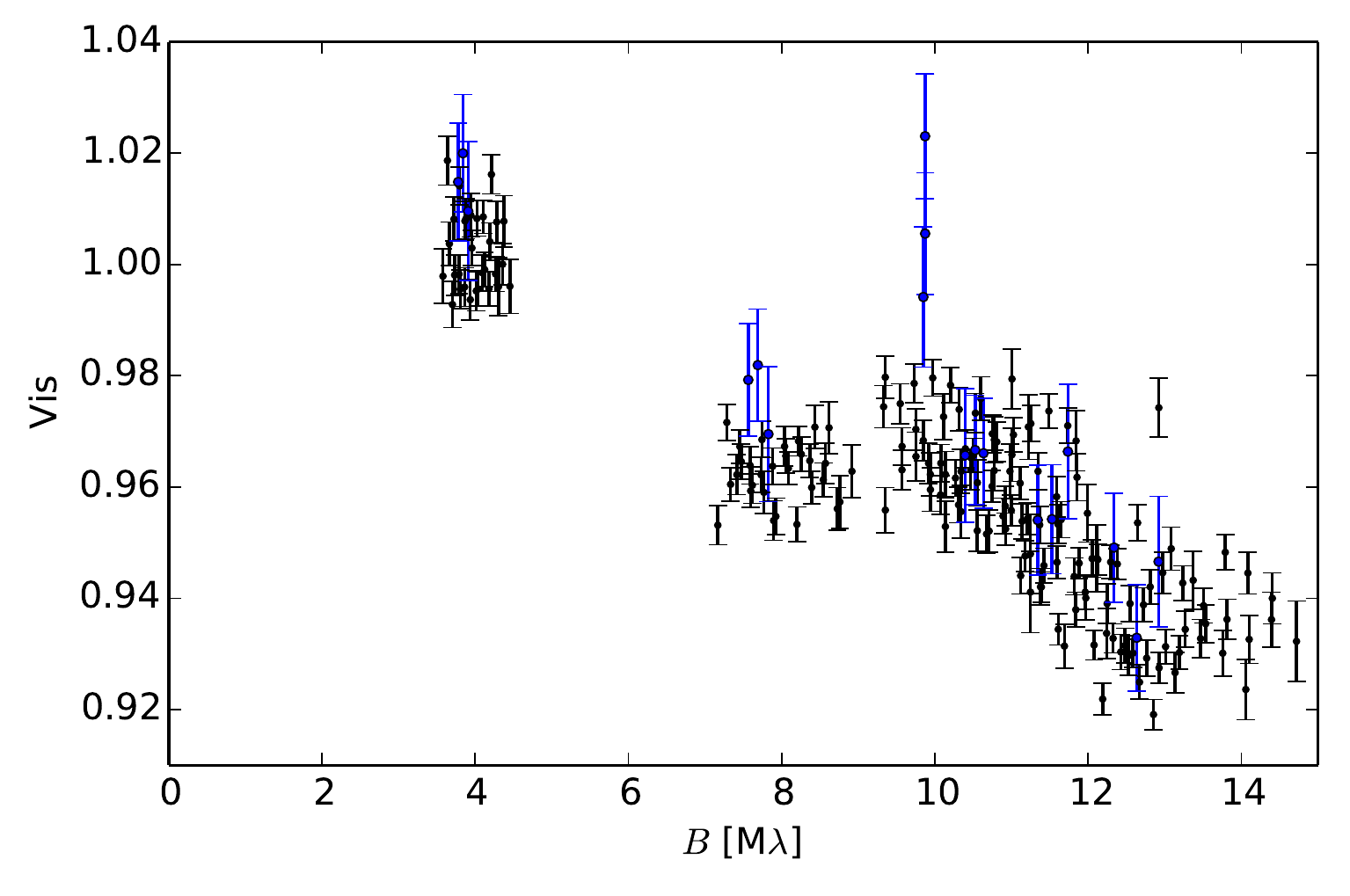}
   \includegraphics[width=0.45\textwidth]{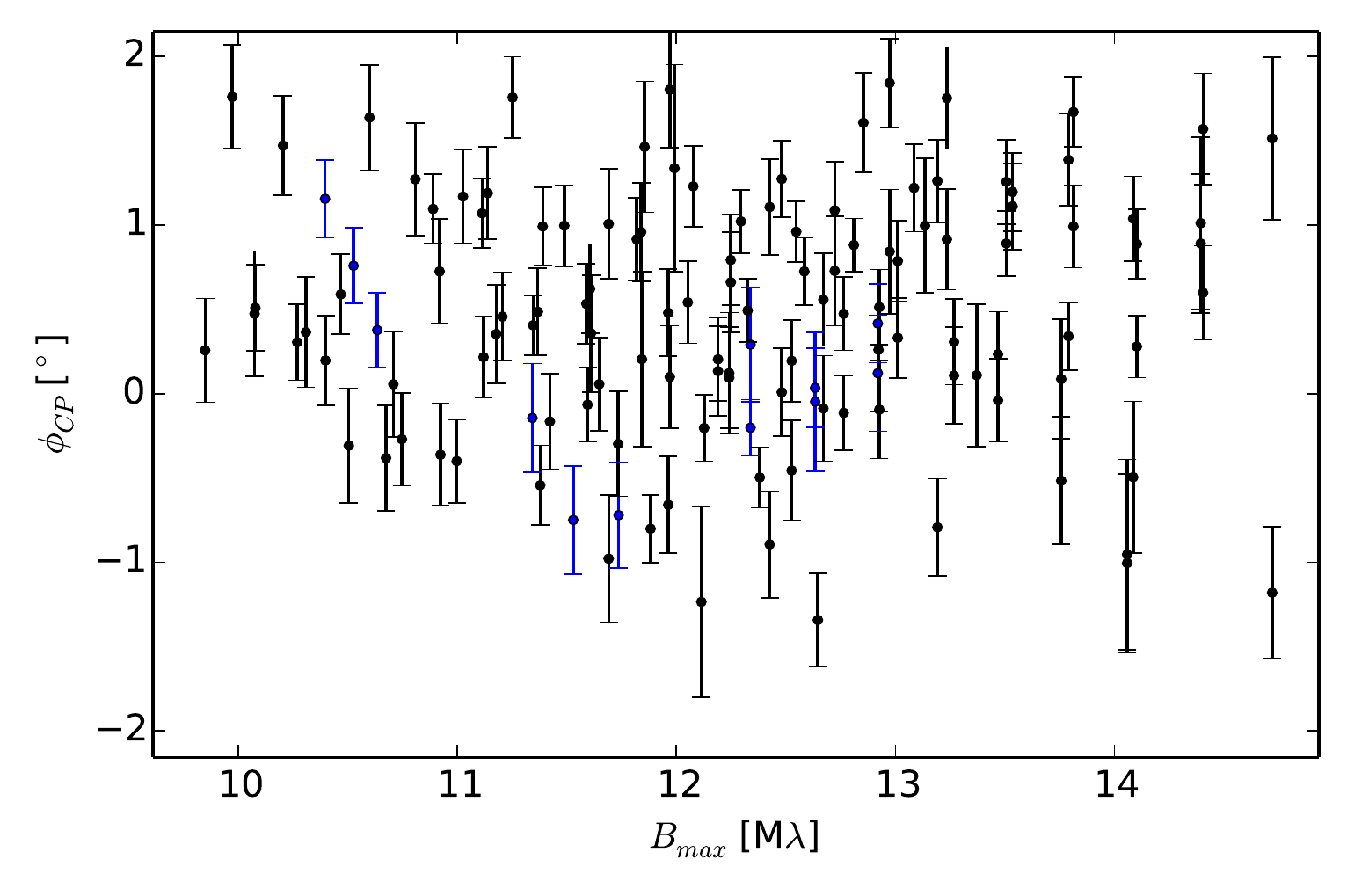}
   \caption{ Visibilities (top) and closure phases (bottom) for the binned medium spectral resolution, science camera data (black data points) and those from the central wavelength channel of the low resolution, fringe tracker data (blue data points).}
  \label{fig:FTvsSC}
\end{figure}

In addition to the science camera data, GRAVITY provides data from the fringe tracker channel. This is always dispersed at low spectral resolution, providing five wavelength channels across the $K$-band. In the analysis presented here, we focused on the medium spectral resolution data obtained via the science camera channel. This was binned to increase the signal-to-noise, resulting in ten wavelength channels across the $K$-band. In Fig.~\ref{fig:FTvsSC}, we compare the binned medium spectral resolution, science camera visibilities and closure phases (shown in black) with the corresponding central wavelength channel values from the low resolution, fringe tracker data (shown in blue), both obtained on UT data 2016-09-11. The data are broadly consistent.

%%%%%%%%%%%%%%%%%%%%%%%%%%%%%%%%%%%%%%%%%%%%%%%%%%

% Don't change these lines
\bsp	% typesetting comment
\label{lastpage}
\end{document}